# Microwave Cherenkov Radiation from a Particle-in-flight to a Semi-infinite Layered Medium

Levon Sh. Grigoryan [a], Alpik R. Mkrtchyan [a], Hrant F. Khachatryan [a], Svetlana R. Arzumanyan [a], and Wolfgang Wagner [b]

[a] Institute of Applied Problems in Physics, Yerevan, 0014 Armenia

[b] Forschungszentrum Dresden-Rossendorf, Institute of Radiation Physics, POB 510119, 01328 Dresden, Germany

*Abstract*—Some part of the microwave Cherenkov radiation from a particle-in-flight from vacuum to semi-infinite layered medium is redirected by the periodical structure of medium in the backward direction. This part of radiation proves to be quasi-monochromatic.

## I. INTRODUCTION AND BACKGROUND

THE presence of matter may essentially influence the characteristics of high energy electromagnetic processes giving rise to the production of Cherenkov radiation, transition radiation etc. The effects of interest arise in periodical structures of different configurations (see, e.g., [1,2] and references therein).

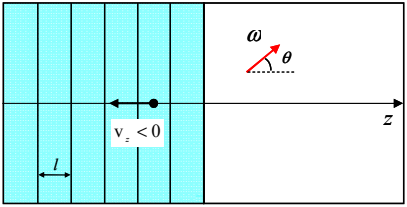

Fig.1. The particle flying from vacuum into the layered medium. $l$ is the period of layered medium

The present paper deals with the radiation from a particle-in-flight from vacuum to a semi-infinite layered medium (see Fig.1, where the particle velocity $v_z = -v < 0$). The features of spectral-angular distribution $I_{v_z<0}(\omega,\theta)$ of the energy

$$E = \int I_{v_z<0}(\omega,\theta) d\omega d\theta \quad (1)$$

of radiation propagating in vacuum in the backward direction (with respect to the direction of particle motion) during the whole time of particle motion have been investigated.

If the sense of particle motion is reversed ($v_z = v > 0$ in Fig.1), one comes to the problem of radiation from a particle flying out of the layered medium to vacuum. In this case $I_{v_z>0}(\omega,\theta)$ describes the propagation of radiation in vacuum in the forward direction relative to the sense of particle motion (radiation in the forward hemisphere).

Below we shall confine to the consideration of radiation with wavelength of the order of the period of layered medium assuming that $l$ ranges from a fraction of a micron to millimeter. In this wavelength range the permittivity and permeability of the layered medium (e.g., photonic crystal) may change in wide limits. Respectively, the effect of the periodicity of layered medium on the radiation from relativistic particle may prove large. However, the analysis of angular distribution of radiation energy is complicated by the fact that propagating inside the layered medium are the Bloch waves (traveling waves modulated with the periodicity of medium), not the plane ones. For this reason we consider the radiation in vacuum outside the borders of the layered medium.

The expressions required for calculation of $I_{v_z}(\omega,\theta)$ were derived in [3]. It was shown ibidem that at the flight of particle from vacuum to the layered medium, some part of transition radiation (TR) emitted in the forward direction [4,5] was redirected by the periodical structure of layered medium in the backward direction, - to vacuum. In this connection in [3] the results of numerical calculations of $I_{v_z<0}(\omega,\theta)$ were given for angles $0 \leq \theta \leq 5°$. In the present work a similar phenomenon was investigated for the case of Cherenkov radiation (CR) from a relativistic particle.

## II. RESULTS

The results of numerical calculations of $I_{v_z}(\omega,\theta)$ for $0 \leq \theta \leq 90°$ are given in Figs.2,3. Fig.2 corresponds to radiation in the forward hemisphere when the particle leaves the semi-infinite layered medium for vacuum (the case of $v_z > 0$ in Fig.1), and Fig.3 – to radiation to the backward hemisphere when the particle flies from vacuum to the semi-infinite layered medium ($v_z < 0$ in Fig.1). The energy of particle (electron) is 30 MeV, the average permittivity of medium $\bar{\varepsilon} = 1.5$, the variation profile of $\varepsilon$ is the same as that in [3], $\Delta\varepsilon = 0.5$ (modulation depth of $\varepsilon$), the loss-angle tangent $\delta = 0.01$, the permeability is 1. The light-colored areas in the figures correspond to larger values of $I_{v_z}(\omega,\theta)$ (the lighter the area, the larger is the value of $I_{v_z}$). For radiation in the forward direction $\max I_{v_z>0}(\omega,\theta) = 3.48\hbar$, and in the backward direction $\max I_{v_z<0}(\omega,\theta) = 0.45\hbar$.

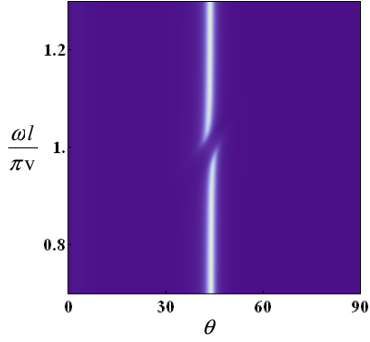

Fig.2. The spectral-angular distribution of radiation energy in the forward direction $I_{v_z>0}(\omega,\theta)$, in vacuum at the departure of particle from semi-infinite layered medium.

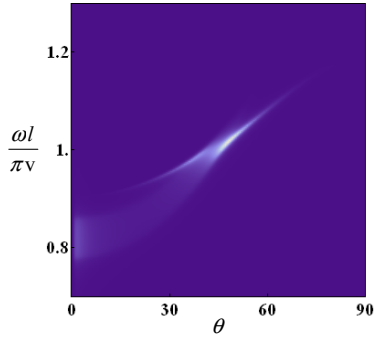

Fig.3. The spectral-angular distribution of radiation energy in the backward direction $I_{v_z<0}(\omega,\theta)$, in vacuum for the electron incoming to semi-infinite layered medium.

As a result of analyzing the data in Figs. 2,3 the following conclusions have been made:

(1) The forward radiation $I_{v_z>0}(\omega,\theta)$ (Fig.2) at the departure of relativistic particle from semi-infinite layered medium to vacuum is directional. Its direction ($\theta = 48,6°$) is sufficiently well determined by the Cherenkov condition and by refraction law on the boundary of semi-infinite layered medium with vacuum if the average value of refractive index is used.

(2) There is a downfall in radiation intensity in the range of $\omega \approx \pi v/l$ frequencies, since the electromagnetic waves with these values of $\omega$ fail to freely propagate in the layered medium. The width of this forbidden band decreases when $\Delta\varepsilon'_1 \to 0$ (and is zero in the limit of semi-infinite uniform medium).

As TR formed at the departure of particle from the semi-infinite medium [4,5] is weak, it is not noticed in Fig.2, whereas the resonance radiation from particle induced by the periodical structure of layered medium [5,6] is emitted beyond the bounds of frequency range shown in Fig.2.

(3) The backward radiation, $I_{v_z<0}(\omega,\theta)$ (Fig.3), in vacuum at the flight of particle into the semi-infinite medium is directional and quasi-monochromatic. The typical direction of radiation is the same as that in Fig.2.

(4) The range of emitted frequencies is adjacent to the forbidden band of the layered medium and is determined by the Bragg diffraction of CR on the periodical structure of this medium. The degree of radiation monochromaticity increases as the modulation depth $\Delta\varepsilon'_1$ decreases.

The contribution of resonance radiation is traced in Fig.3 as thin offshoots branching from the central peak. As is seen in Fig.3, $I_{v_z<0}(\theta)$ reaches the local maximum $= 0.07\hbar$ for $\theta = 1.8°$. The latter corresponds to TR from the relativistic particle diffracted backwards by the periodical structure of the layered medium [3]. Unlike TR, the value of $I_{v_z<0}(\omega,\theta)$ in the area of central peak is unlimitedly increasing when $\delta \to 0$ (increase in the transparency of the material of semi-infinite layered medium).

So, the periodicity of medium has the following effects on the radiation of charged particle:

- In the range of wavelengths of the order of period $l$ of the layered medium the relativistic particle does not generate CR in the forward direction.
- The part of CR spectrum adjacent to this forbidden band is redirected by the periodical structure of medium in the backward direction relative to the sense of particle motion.

Here:

- (Owing to the influence of particle) the layered medium generates CR and simultaneously redirects it back.
- The detection of this radiation permits to determine the period of layered medium by the radiation wavelength.

The periodical structure with tuned $l$ and $\Delta\varepsilon$ may be induced in the medium by using, e.g., ultrasonic vibrations. If that occurs one may control the radiation wavelength in the millimeter and sub-millimeter ranges by attuning ultrasonic vibrations in the range of frequencies in excess of $10 MHz$.